\documentclass[journal=jacsat,manuscript=article]{achemso}

\usepackage[version=3]{mhchem} 
\usepackage{graphicx} 

\author{Ebrahim Forati}
\author{George W. Hanson}
\email{george@uwm.edu}
\affiliation[Unknown University]
{Department of Electrical Engineering and Computer Science, University of Wisconsin Milwaukee, WI53211, USA}

\title[soft-boundary graphene nanoribbons]
{Surface plasmon polaritons on soft-boundary graphene nanoribbons
and their application as voltage controlled plasmonic switches and
frequency demultiplexers}

\begin{document}
\begin{abstract}
  A graphene sheet gated with a ridged ground plane, creating a soft-boundary
(SB) graphene nanoribbon, is considered. By adjusting the ridge parameters
and bias voltage a channel can be created on the graphene which can
guide TM surface plasmon polaritons (SPP). Two types of modes are
found; fundemental and higher-order modes with no apparent cutoff frequency and with energy distributed
over the created channel,  and edge modes with energy concentrated
at the soft-boundary edge. Dispersion curves, electric near-field
patterns, and current distributions of these modes are determined.
Since the location where energy is concentrated in the edge modes
can be easily controlled electronically by the bias voltage and frequency,
the edge-mode phenomena is used to propose a novel voltage controlled
plasmonic switch and a plasmonic frequency demultiplexer.
\end{abstract}


\section{Introduction\label{sec:Introduction} }

Graphene is the first two dimensional atomic crystal available to
researchers, and has been the subject of intense study concerning
both its fabrication and applications \cite{Novoselov,Zhang,Berger,Geim,Nair,Bonaccorso,Schedin,Geim_science,Mak,Mueller,Xia,Lee,Castro}.
Electrical properties of graphene, as represented by a local conductivity,
are studied in many papers \cite{Falkovsky,Falkovsky-1,Mikhailov,Gusynin-1,GusyninP,Peres,Hanson-1,W-Hanson,Peres_1,Ziegler}. It has
been shown that the conductivity of graphene consists of interband
and intraband contributions whose imaginary parts have different signs.
Assuming an $e^{i\omega t}$ time convention, at lower frequencies
and lower temperatures the intraband term with negative imaginary
part dominates the conductivity, otherwise the conductivity has a
positive imaginary part due to the interband contribution. Furthermore,
it has been shown that a graphene sheet can support a single TM surface
plasmon polariton, and only in the regime where the conductivity has
negative imaginary part \cite{Mikhailov,GWHanso}. Likewise,
when conductivity has a positive imaginary part only a TE  mode
can propagate. However, TE modes are very loosely confined to the
graphene surface and are not considered further here. 

\begin{figure}
\includegraphics[scale=0.3]{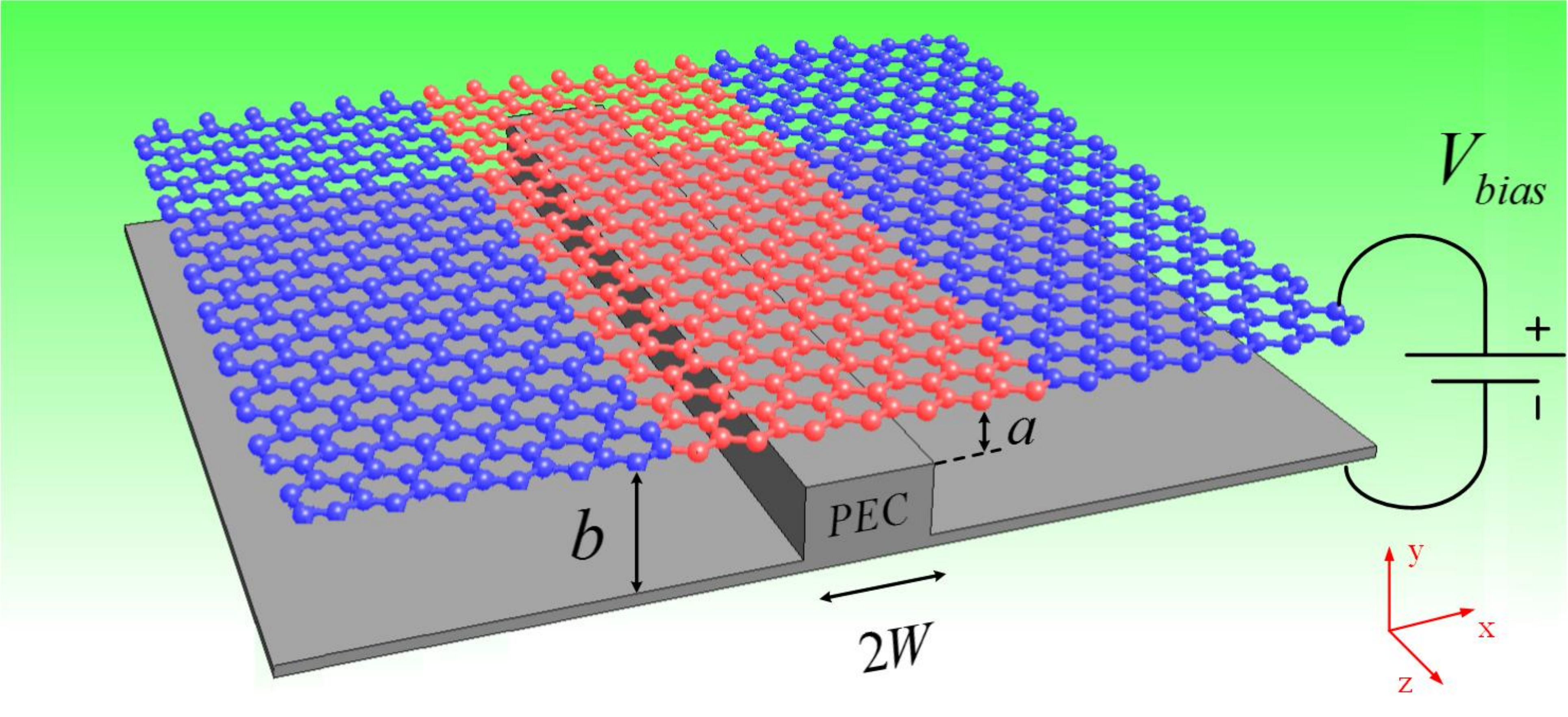}

\caption{Graphene sheet gated with a ridged, perfect electrically-conducting (PEC) ground plane for the electrostatic bias, forming a soft-boundary graphene nanoribbon. The red area depicts the SPP
channel having $\mathrm{Im\left(\sigma\left(x\right)\right)<0}$,
and the blue area depicts the region where $\mathrm{Im\left(\sigma\left(x\right)\right)>0}$
and SPP propagation is prohibited. }
\end{figure}

The conductivity of graphene can be controlled by its carrier density,
which can be varied by an electrostatic or magnetostatic bias, and/or
chemical doping. This fact is used in \cite{Switch} to implement a graphene based plosmonic switch. Recently, it has been proposed to use a perturbed
ground plane as depicted in Fig. 1 in order to obtain different conductivities
on the graphene, near and far from the ridge, by the use of a single
bias voltage \cite{Vakil}, thereby creating a SPP propagation channel
parallel to the ridge. In \cite{Vakil} the ground plane ridge was
assumed to form a conductivity profile with sharp features, i.e.,
to effectively form a hard-boundary (HB) graphene nanoribbon (GNR)
wherein $\mathrm{Im}\left(\sigma\right)<0$ for $\left|x\right|<W$
and $\mathrm{Im}\left(\sigma\right)>0$ for $\left|x\right|>W$, where
$2W\mbox{ is the ridge width}$ and $\sigma$ is assumed to be constant
in each region. To remove the HB assumption, in \cite{IOP} we studied
the role of geometry and bias on forming the desired channel. In that
work, rather than assuming a piece-wise constant conductivity, we
determined the actual conductivity profile $\sigma\left(x\right)$
by finding the electrostatic charge distribution $\rho\left(x\right)$
on the graphene sheet from Laplace's equation, leading to the chemical
potential $\mu_{\mathrm{c}}\left(x\right)$ and the conductivity via the Kubo
formula. It was shown that the ridged structure does indeed allow
for the formation of a channel in the vicinity of the ridge for
SPP propagation using a single bias, but that the resulting boundary
has, as expected, a softened profile (i.e., a soft boundary (SB))
wherein the conductivity is not constant. The work \cite{IOP} was
concerned with the properties of the soft boundary and resulting channel,
and the current distribution of the fundamental SPP mode. In this
work we consider the various other modes that can propagate along
the SB channel, including higher-order modes and edge modes. In particular,
we show that unlike the HB case, for a soft boundary the higher-order
modes have no apparent low-frequency/long-wavelength cutoff, although
as frequency is lowered modal energy tends to spread out laterally
along the effectively wider channel. We also show that low-loss edge
modes can propagate for which the location where energy is concentrated
can be controlled electronically. We then consider two applications
of the structure, as a plasmonic voltage-controlled switch and a frequency
demultiplexer. 

\begin{figure}
\begin{centering}
\includegraphics[scale=0.3]{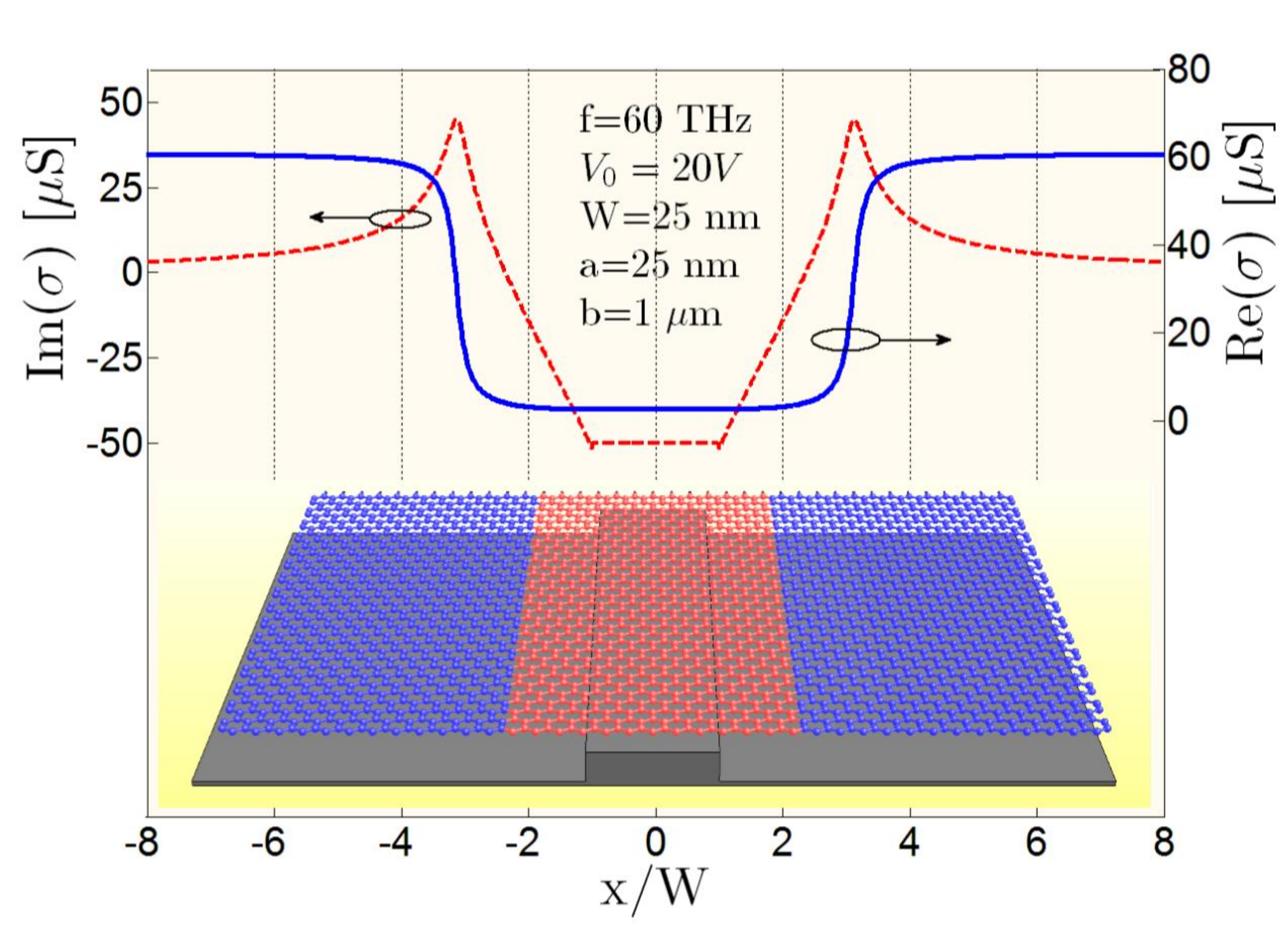}
\par\end{centering}

\caption{Conductivity distribution on the graphene sheet. }
\end{figure}

Figure 2 shows the conductivity profile $\sigma\left(x\right)$ of
the graphene sheet for a representative set of geometrical and electrical
parameters; the SPP channel terminates where $\mathrm{Im}\left(\sigma\right)$
becomes positive. The slope of the conductivity in the vicinity of
the soft boundary can be adjusted by the ridge parameters, discussed
in detail in \cite{IOP}.

Throughout this work, the parameters of Fig. 1 are set as $a=25\mathrm{nm}$,
$b=1\mu\mathrm{m}$, $V_{0}=20\mathrm{V},$ and $W=25\mathrm{nm}$.
The mathematical formulations and numerical procedure are given in
the supporting information\textbf{.}

\section{Soft boundary SPP modes }

\begin{figure}[tbh]
\begin{centering}
\includegraphics[scale=0.25]{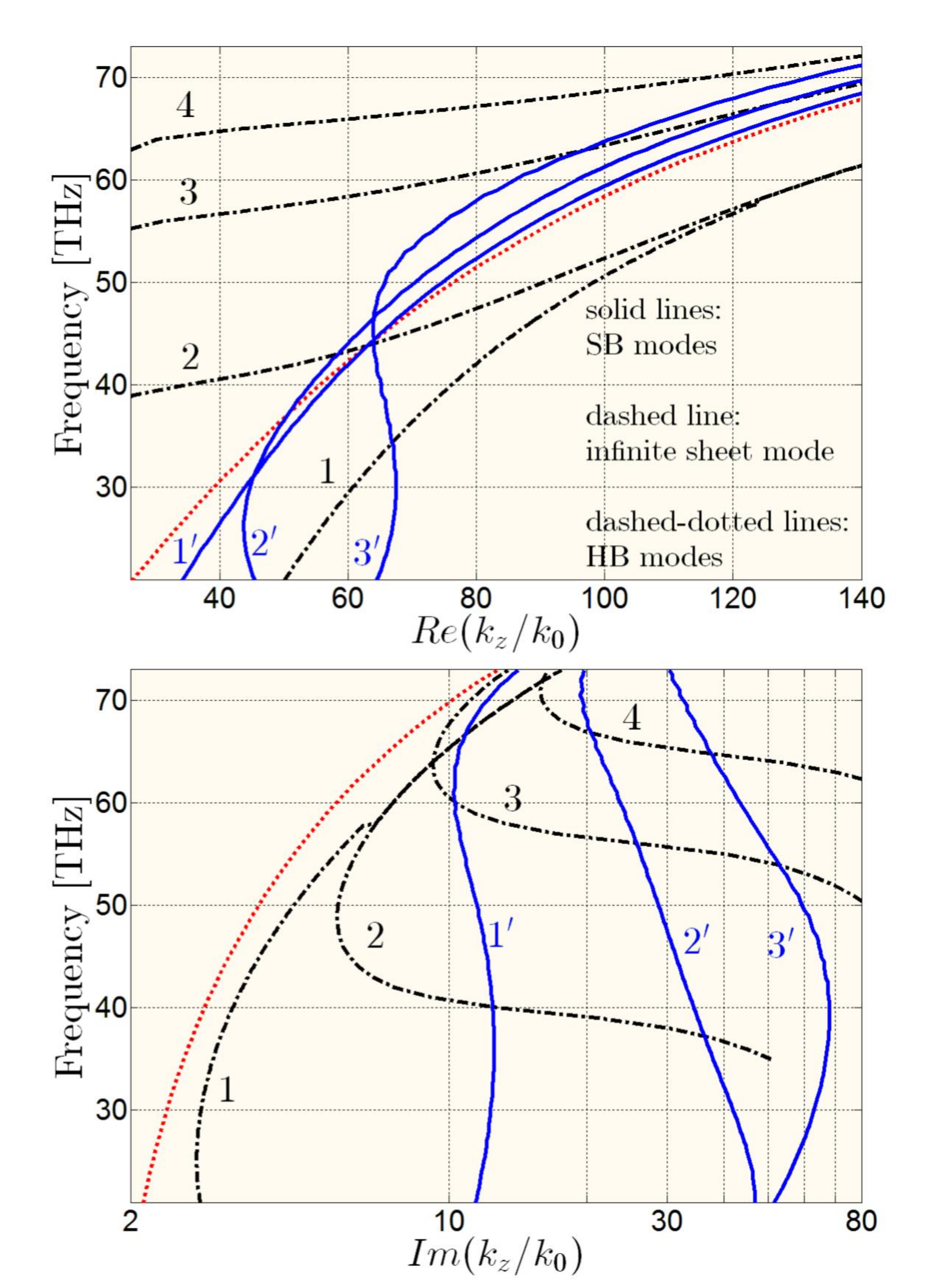}
\par\end{centering}

\caption{Modal dispersion curves for the SB case (blue). For convenience, the SPP
for an infinite graphene sheet (at $\mu_{\mathrm{c}}=0.239\:\mathrm{eV}$)
is also shown (red), as well as the modes of a HB GNR (black) with the same conductivity
as the SB case for $\left|x\right|<W$ (and $\sigma=0$ for $\left|x\right|>W$). }
\end{figure}

\begin{figure}[!t]
\includegraphics[scale=0.16]{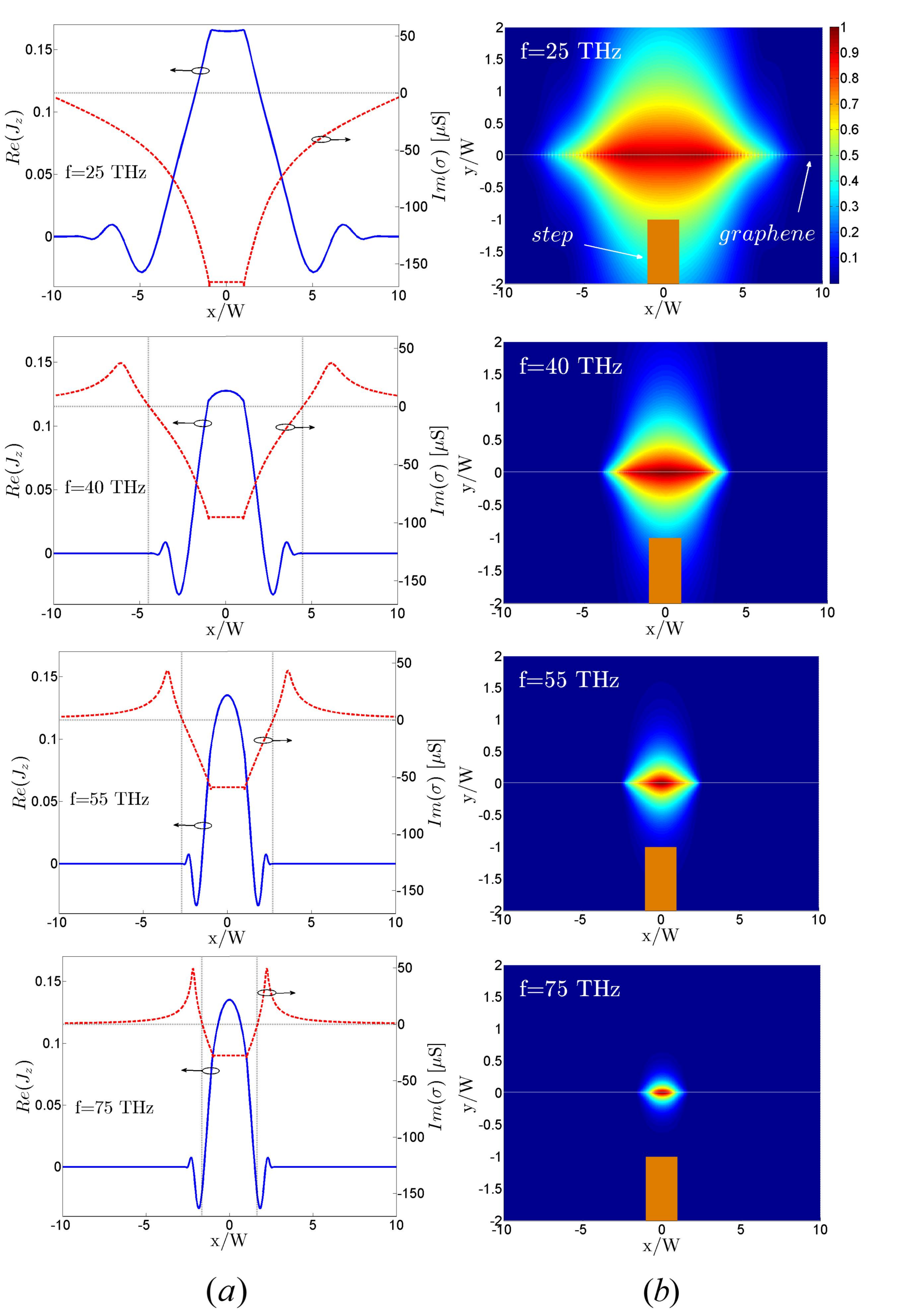}
\caption{$\mathrm{Re}\left(J_{\mathrm{z}}\left(x\right)\right)$ and $\left|E_{\mathrm{z}}\left(x\right)\right|$distributions of the first SB mode ($1^{\prime}$ in Fig. 3) at four different frequencies. The conductivity profile $\left(\mathrm{Im}\left(\sigma\right)\right)$ is also shown.}
\end{figure}

SPP modes for the geometry in Fig. 1 are calculated by a rigorous
electromagnetic analysis based on an electric-field integral equation
(see \cite{IOP} and supporting information). Figure 3 shows the dispersion
curves of the first three soft boundary modes.
Also shown are the first four modes of a suspended%
\footnote{We consider a suspended graphene nanoribbon rather than a nanoribbon
over a ground plane since the modes are strongly confined to the graphene
surface and do not interact with the ground plane (which merely serves
to bias the graphene sheet). %
} graphene nanoribbon with width $50\mathrm{nm}$ and the same conductivity
as for the SB modes for $\left|x\right|<W$, with $\sigma=0$ for
$\left|x\right|>W$. These modes are marked as hard boundary (HB)
modes, and are the usual GNR modes. As discussed in several papers
\cite{Nikitin,Sounas}, for the HB GNR the first two modes (called
edge modes since energy is concentrated at the graphene edge, similar
to that found for a graphene half-space) asymptotically merge to the
edge mode of a semi-infinite graphene sheet \cite{Sounas,Apell}.
The other higher order modes of the HB GNR become asymptotic to the
single TM SPP mode \cite{W-Hanson} that can propagate on an infinite graphene sheet. This HB behavior is shown in Fig. 3, where we also show that, unlike the HB case, for the soft boundary all modes become asymptotic to the infinite sheet SPP. In this regard, SB modes
of the geometry are analogous to the sufficiently-higher-order modes (modes 3,4,
...) of the HB GNR. 

Figure 3 shows that there is no apparent cut off frequency for the
SB modes. This is because, unlike a HB GNR with fixed width, as frequency
decreases the effective width of the SB channel increases (i.e., the
effect of the ridge perturbation which creates the propagation channel
on the graphene sheet extends further away from the ridge as frequency
is lowered). To clarify this, Fig. 4 shows the current, conductivity
and field distribution of the first SB mode. Fig. 4(a) shows the normalized
real part of the longitudinal current, $\mathrm{Re}\left(J_{\mathrm{z}}\right)$,
and the imaginary part of the conductivity, $\mathrm{Im}\left(\sigma\right)$,
as a function of $x$. Since these are eigenmode currents, they are
normalized so that $\left\Vert J_{\mathrm{T}}\right\Vert =1$, where $J_{\mathrm{T}}$
is the total current consisting of longitudinal and transverse components
(i.e. $\int\left(\left|J_{\mathrm{x}}\left(x\right)\right|^{2}+\left|J_{\mathrm{z}}\left(x\right)\right|^{2}\right)dx=1$).
Fig. 4(b) shows the normalized magnitude of the longitudinal electric
field $\left|E_{\mathrm{z}}\right|$ in the transverse coordinate ($x-y$).
Here we define an effective width for the channel ($W_{\mathrm{eff}}$) as
the width of the channel on the graphene sheet where $\mathrm{Im}\left(\sigma\left(x\right)\right)<0$,
which occurs in the vicinity of the ridge (since only in this region
can SPPs propagate). It is evident from Fig. 4 that $W_{\mathrm{eff}}$ increases
as frequency decreases and therefore there is no cut off frequency
for SB modes as exists for HB GNR modes (and, more generally, for
all waveguides of fixed transverse dimensions), which is an important
distinction between the SB and HB cases. Quantitatively, $W_{\mathrm{eff}}/W$
is 21.13, 8.97, 5.4, and 3.3 at 25, 40, 55, and 75 THz, respectively.

The current distribution for SB modes is similar to those of HB GNR
modes in the region $\left|x\right|<W$. Outside of this region the
current vanishes after a few oscillations. These oscillations resemble
the field distribution in the cladding of an optical fiber with graded
index cladding \cite{Kong}. Since $\mathrm{Im}\left(\sigma\right)>0$
in the region $\left|x\right|>W_{\mathrm{eff}}$ , the current (and therefore
the mode) is forced to be limited to the $\left|x\right|<W_{\mathrm{eff}}$
region. However, this is not a necessary condition for mode confinement
(although it is sufficient). In \cite{IOP} we show that even when
$\mathrm{Im}\left(\sigma\right)<0$ everywhere on the graphene sheet
(e.g., at low frequency and higher temperature) it is still possible
to have modes confined near the ridge. In fact, the confinement condition
is that the conductivity boundaries are sharp enough so that the current concentrates
in the vicinity of the ridge. As the boundaries become softer, by,
say, lowering the ridge, the currents spread out further. 
\begin{figure}
\centering{}\includegraphics[scale=0.25]{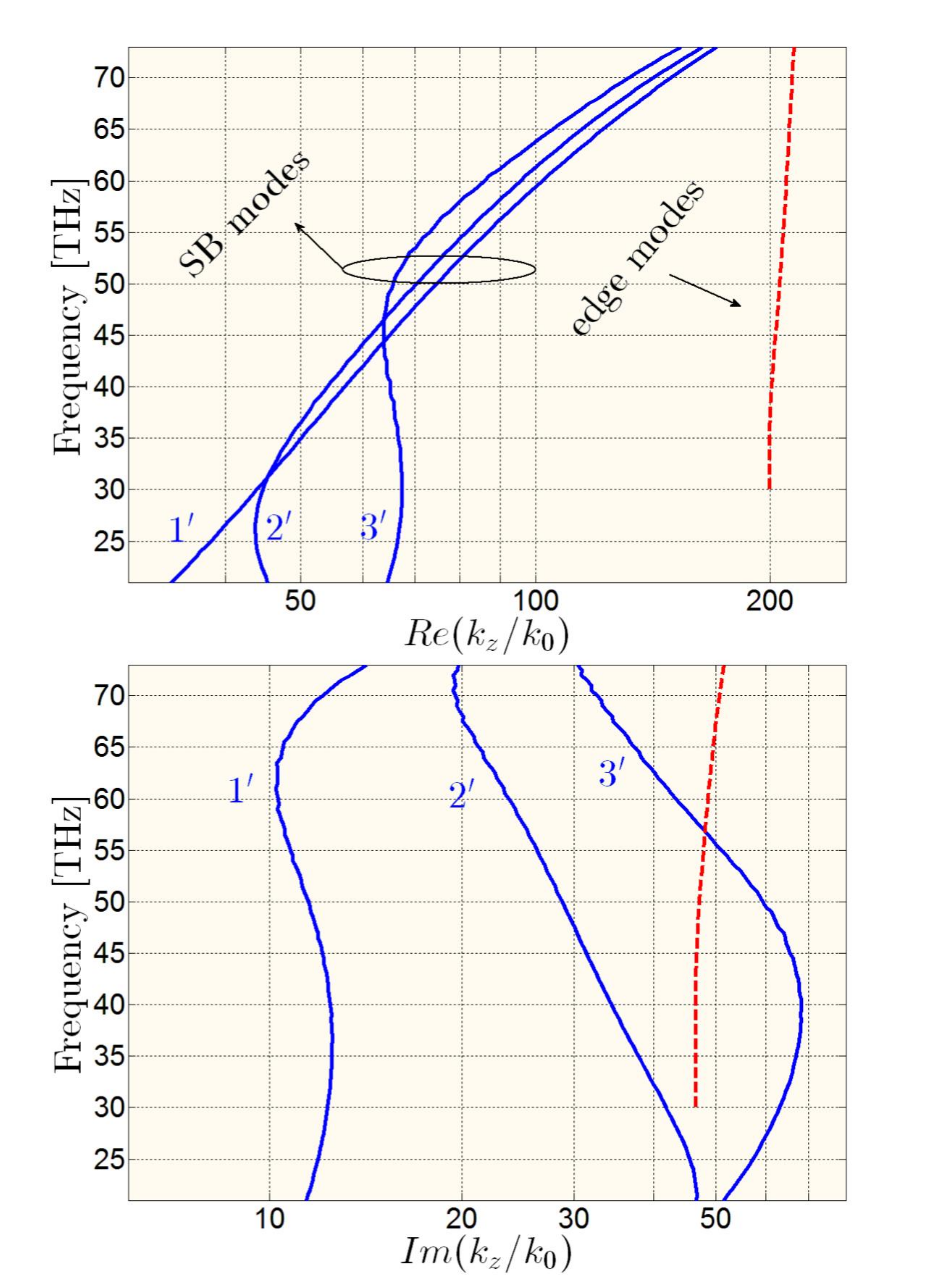}
\caption{Dispersion curves of the geometry of Fig. 1 showing the previously-discussed bulk-like SB modes, and SB edge modes.}
\end{figure}

The current and field distributions for the second and third SB modes
are provided in the supporting information. Further higher order SB modes
are not as important since they are very lossy.

\section{Soft boundary edge modes }

\begin{figure}[thb]
\includegraphics[scale=0.16]{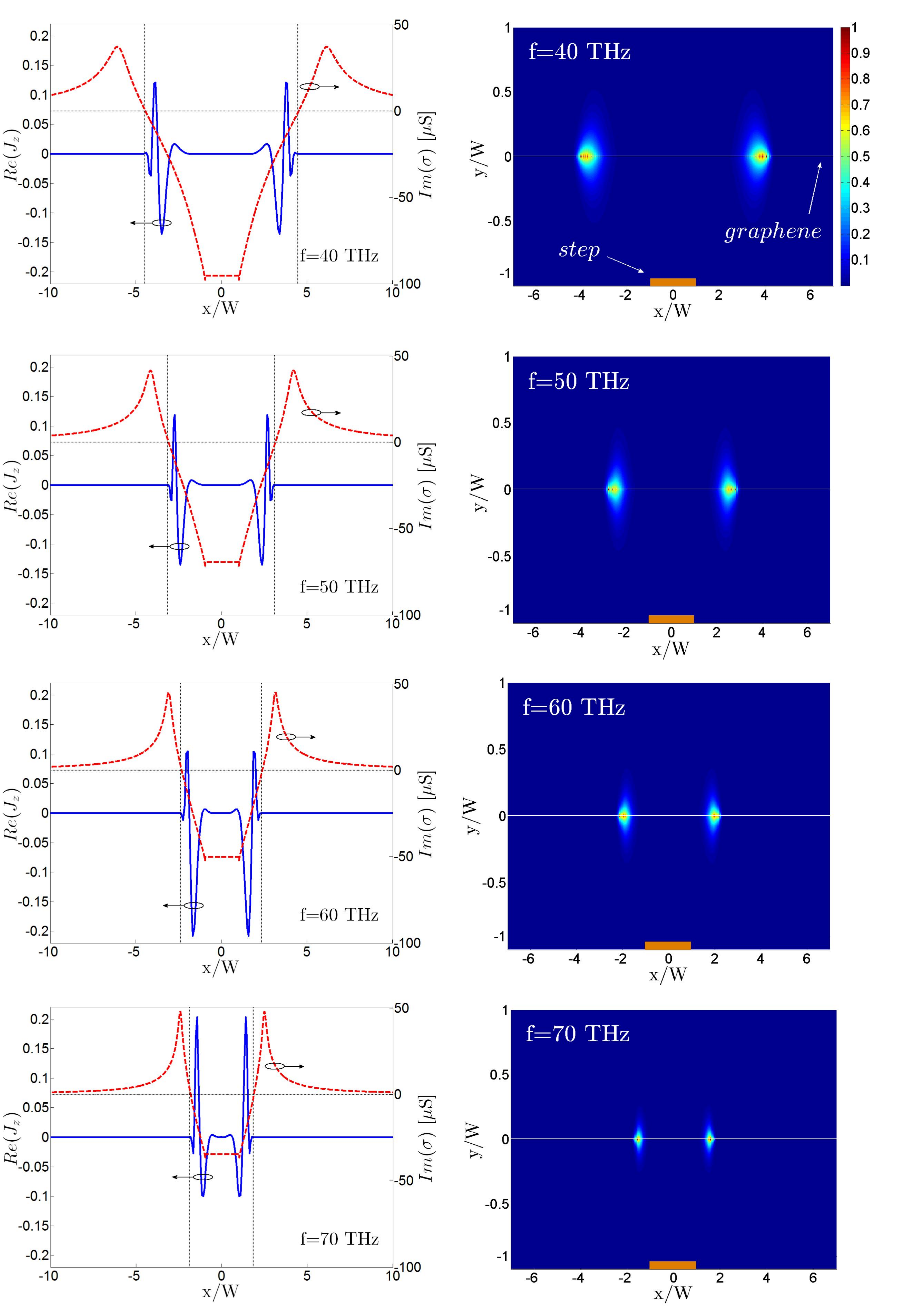}
\caption{$\mathrm{Re}\left(J_{\mathrm{z}}\left(x\right)\right)$ and $\left|E_{\mathrm{z}}\left(x\right)\right|$distributions
of the even edge mode at four different frequencies.}
\end{figure}

As with a HB GNR, there are two degenerate edge modes for the geometry
in Fig. 1, as considered in Fig. 5. These two even and odd edge modes
propagate along edges of the created channel (i.e. in the region where
$\mathrm{Im}\left(\sigma\right)$ changes sign). As Fig. 5 shows,
edge modes are slower and less dispersive than the previously-discussed SB bulk-like (since current spreads out over the bulk of the created channel) modes. Therefore,
at lower frequencies where the boundary is softer, these modes become
more important since they are much slower than other SB modes. Fig.
6 shows the current and field distributions of the even edge mode
for four different frequencies. Obviously, the even and odd classification
can be interchanged by the left and right edge modes as is done in
some works \cite{Nikitin,Sounas}. The SB edge modes have
two important properties: First, they are fairly low-loss and slow,
and therefore more tightly confined to the graphene surface then the
bulk SB modes (which, nevertheless, are fairly tightly confined to
the graphene surface). Second, their physical location relative to
the ridge varies with the applied bias voltage and with frequency.
The latter property makes the geometry useful for switching and demultiplexing
applications, as proposed in the next section.

\section{Plasmonic switch and demultiplexer}

\begin{figure}[tbh]
\includegraphics[scale=0.25]{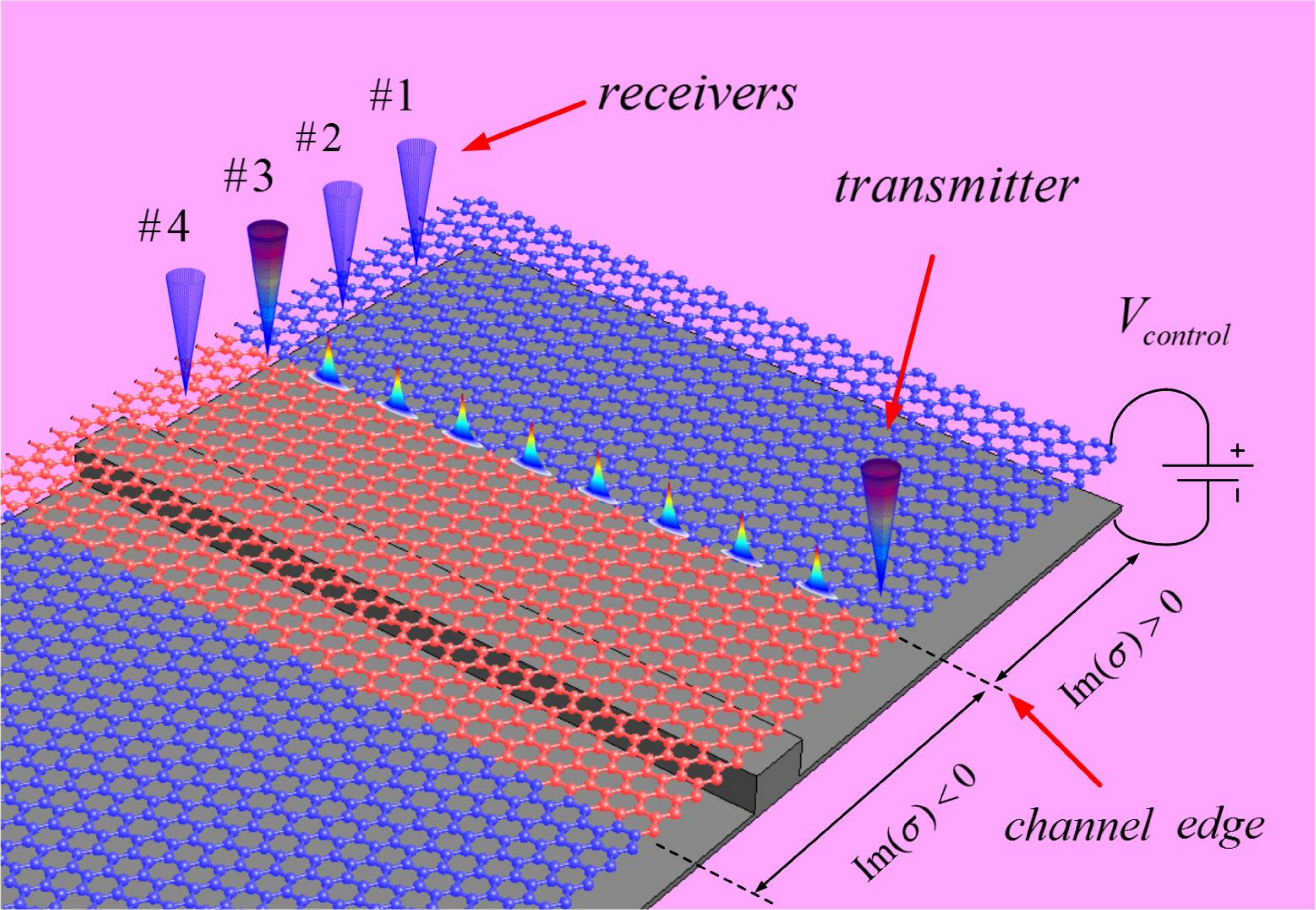}

\caption{A scheme to form a voltage-controlled plasmonic switch/demultiplexer.}
\end{figure}

Since the effective width of the channel can be controlled by the
bias voltage, the physical location where energy is concentrated for
the edge modes can be easily controlled. One of the applications of
this phenomena is a plasmonic voltage controlled switch as depicted
in Fig. 7. By adjusting $V_{\mathrm{control}}$ in Fig. 7, the edge
of the channel can be aligned with one of the receivers, and the edge
mode can transfer energy from the transmitter to the desired receiver.
In Fig. 7, the ridge parameters and bias voltage are assumed to form
the channel such that it's edge is aligned with receiver 3. 

On the other hand, it is evident from Fig. 6 that the channel width
varies as frequency changes even with a fixed bias voltage. This suggests
that we can design a plasmonic demultiplexer with the same geometry
as shown in Fig. 7 with a fixed $V_{\mathrm{control}}$. Then, e.g.,
lower frequencies can be transferred to receiver $\#1$ and higher
frequencies can be guided to, e.g., receiver $\#4$. The characteristics
of these plasmonic switches and demultiplexers can be determined by
adjusting the distance between the receivers, ridge parameters, and
the bias voltage range. Of course, it is assumed that the transmitter
can excite a fairly wide area on the graphene sheet relative to the
receivers, receivers, to allow the edge mode to be excited by a fixed-location transmitter.

\section{Conclusion}

Different SPP modes have been studied for the geometry depicted in
Fig. 1, forming a soft boundary graphene nanoribbon. The dispersion
curves, current, and field distributions were numerically calculated
for a few bulk SB modes and for SB edge modes. It was shown that,
unlike a hard boundary graphene nanoribbon, the bulk modes of the
geometry do not have cut off frequencies due to the fact that the
effective width of the channel increases as frequency decreases. It
was also observed that the position of the edge modes varies according
to the bias voltage and frequency. A novel voltage controlled plasmonic
switch and plasmonic frequency multiplexer were proposed utilizing
the SB edge modes, which propagate with low loss. 

\suppinfo

The carrier density distribution of Fig. 1 (the graphene sheet is in the $y=0$ plane  and the center of the ridge is at $x=0$) is \cite{IOP}

\begin{equation}
\frac{\rho\left(x\right)}{\varepsilon_{0}V_{0}}=\begin{cases}
\begin{array}{c}
\frac{1}{a}\\
\frac{1}{b}+{\displaystyle \sum_{n=1}^{\infty}}\frac{n\pi}{b}C_{\mathrm{n}}\left(-1\right)^{n}e^{-\frac{n\pi}{b}\left(\left|x\right|-W\right)}
\end{array} & \begin{array}{c}
\left|x\right|<W\\
\left|x\right|>W
\end{array}\end{cases}\label{eq:rho}
\end{equation}
where

\begin{equation}
C_{\mathrm{n}}=-\frac{2b}{a}\left(\frac{1}{n\pi}\right)^{2}sin\left(n\pi\left(1-\frac{a}{b}\right)\right).\label{eq:Cn}
\end{equation}

In obtaining (2),  a zeroth order approximation has been used to assume an x- independent potential in the region above the step ($\left|x\right|<W$). Otherwise, the problem needs to be solved numerically (e.g., by expanding the potentials as series for both $\left|x\right|<W$ and $\left|x\right|>W$ regions). The zeroth-order solution is a good approximation for $W\ll b$ and/or $a\ll b$ in Fig. 1.

This leads to the chemical potential 

\begin{equation}
\mu_{\mathrm{c}}\left(x\right)=\frac{\hbar}{e}v_{\mathrm{F}}\sqrt{\frac{\pi\rho\left(x\right)}{e}}\label{eq:chemical pot}
\end{equation}
where $v_{\mathrm{F}}=9.546\times10^{5}$ m/s is the Fermi velocity. The chemical
potential is then used in the Kubo formula to find the graphene conductivity
distribution $\sigma\left(x\right)$, \cite{Gusynin}

\[
\sigma\left(x\right)=\frac{je^{2}}{\pi\hbar^{2}\left(\omega-j\Gamma\right)}\intop_{0}^{\infty}\varepsilon\left(\frac{\partial f_{\mathrm{d}}\left(\varepsilon,x\right)}{\partial\varepsilon}-\frac{\partial f_{\mathrm{d}}\left(-\varepsilon,x\right)}{\partial\varepsilon}\right)d\varepsilon\qquad\qquad\qquad\qquad
\]

\begin{equation}
\qquad\qquad\,-\frac{je^{2}\left(\omega-j\Gamma\right)}{\pi\hbar^{2}}\intop_{0}^{\infty}\frac{f_{\mathrm{d}}\left(-\varepsilon,x\right)-f_{\mathrm{d}}\left(\varepsilon,x\right)}{\left(\omega-j\Gamma\right)^{2}-4\left(\varepsilon/\hbar\right)^{2}}d\varepsilon,\label{eq:cond}
\end{equation}
where $-e$ is the charge of an electron, $\hbar$ is the reduced Plank's
constant, $f_{\mathrm{d}}\left(\varepsilon,x\right)=\left(exp\left(\frac{\varepsilon-\mu_{\mathrm{c}}\left(x\right)}{k_{\mathrm{B}}T}\right)+1\right)^{-1}$is
the Fermi-Dirac distribution, $k_{\mathrm{B}}$ is the Boltzmann's constant,
and $\Gamma=10^{13}$ 1/s is the phenomenological scattering rate. 

The eigenmodes of the structure are found starting with Ohm's
law

\begin{equation}
\mathbf{J}\left(x,\beta_{\mathrm{z}}\right)=\sigma\left(x\right)\mathbf{E}\left(x,b,\beta_{\mathrm{z}}\right),\label{eq:Ohms}
\end{equation}
where the Fourier transform pair is defined as

\begin{equation}
\mathbf{E}\left(x,y,\beta_{\mathrm{z}}\right)=\intop_{-\infty}^{\infty}\mathbf{E}\left(x,y,z\right)e^{-j\beta_{\mathrm{z}}z}dz
\end{equation}

\begin{equation}
\mathbf{E}\left(x,y,z\right)=\frac{1}{2\pi}\intop_{-\infty}^{\infty}\mathbf{E}\left(x,y,\beta_{\mathrm{z}}\right)e^{j\beta_{\mathrm{z}}z}d\beta_{\mathrm{z}}.
\end{equation}
Current and electric field are related as 

\begin{equation}
\mathbf{E}\left(x,y,\beta_{\mathrm{z}}\right)=\left(k_{0}^{2}+\nabla_{\beta_{\mathrm{z}}}\nabla_{\beta_{\mathrm{z}}}.\right)\qquad\qquad\qquad\label{eq:elec}
\end{equation}

\[
\qquad\qquad\qquad\intop_{x^{\prime}}g\left(x,y,x^{\prime},\beta_{\mathrm{z}}\right)\frac{\mathbf{J}\left(x^{\prime},\beta_{\mathrm{z}}\right)}{j\omega\varepsilon_{0}}dx^{\prime}
\]
where $\nabla_{\beta_{\mathrm{z}}}=\frac{d}{dx}\hat{\mathbf{x}}+\frac{d}{dy}\hat{\mathbf{y}}+j\beta_{\mathrm{z}}\hat{\mathbf{x}}$
and the Green's function is

\begin{equation}
g\left(x,y,x^{\prime},\beta_{\mathrm{z}}\right)=\frac{1}{2\pi}K_{0}\left(\sqrt{\beta_{\mathrm{z}}^{2}-k_{0}^{2}}\sqrt{\left(x-x^{\prime}\right)^{2}+\left(y-b\right)^{2}}\right),\label{eq:Greens}
\end{equation}
$K_{0}\left(x\right)$ being the zero order modified Bessel function
of the first kind. 

Equations (5) and (8) form an integral equation
whose null space gives the eigenmodes of the structure (i.e., different
$\beta_{\mathrm{z}}$ and their associated currents). The pulse function collocation method is used to solve the integral equation, with point matching at the center of the pulses. The eigencurrents in Figs. 4 and 6 are
normalized so that the 2-norm of the eigen current vector (consists
of transverse and longitudinal components) is unity ($\int\left(\left|J_{\mathrm{x}}\left(x\right)\right|^{2}+\left|J_{\mathrm{z}}\left(x\right)\right|^{2}\right)dx=1.$)
\begin{figure}
\begin{centering}
\includegraphics[scale=0.16]{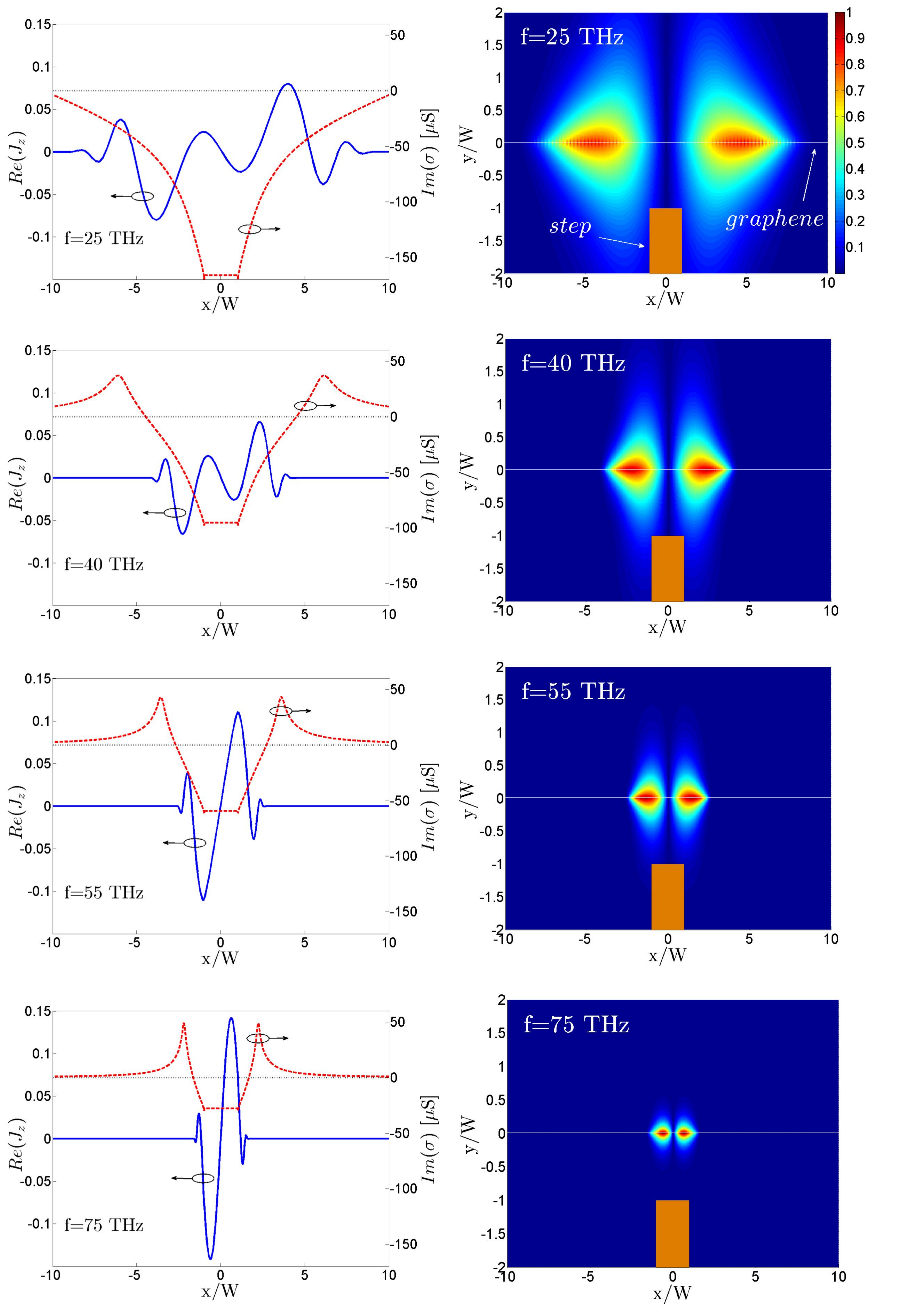}
\par\end{centering}

\caption{$\mathrm{Re}\left(J_{\mathrm{z}}\left(x\right)\right)$ and $\left|E_{\mathrm{z}}\left(x\right)\right|$distributions
of the second SB mode ($2^{\prime}$ in Fig. 3) at four different
frequencies.}
\end{figure}

\begin{figure}
\begin{centering}
\includegraphics[scale=0.16]{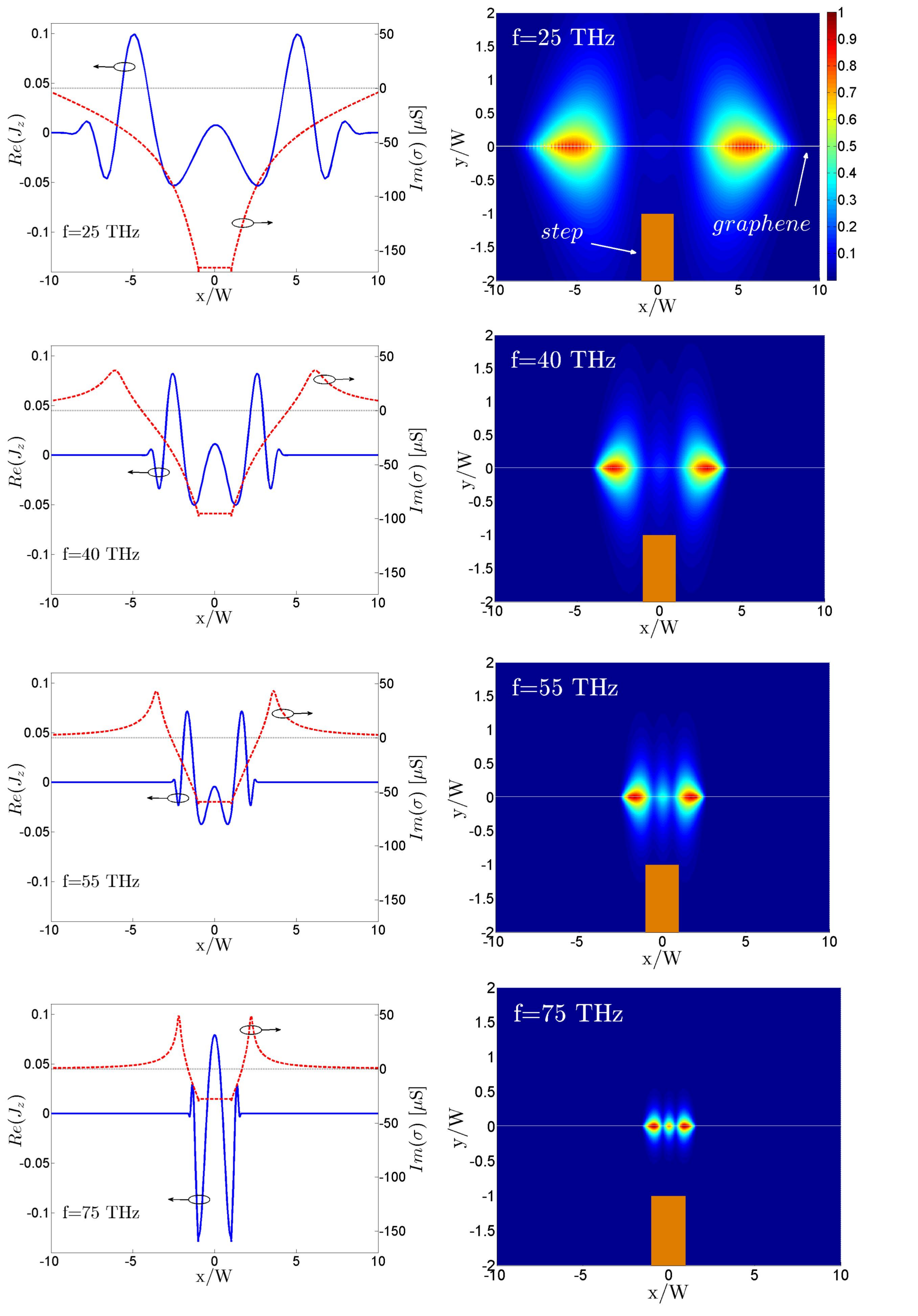}
\par\end{centering}

\caption{$\mathrm{Re}\left(J_{\mathrm{z}}\left(x\right)\right)$ and $\left|E_{\mathrm{z}}\left(x\right)\right|$distributions
of the third SB mode ($3^{\prime}$ in Fig. 3) at four different frequencies.}
\end{figure}

After finding the currents associated with the modes, (8)
is used to find the field distributions 

\begin{singlespace}
\begin{equation}
E_{\mathrm{x}}\left(x,y\right)=\frac{k_{0}^{2}}{2\pi j\omega\varepsilon_{0}}\intop_{x^{\prime}}K_{0}\left(\alpha\right)J_{\mathrm{x}}\left(x^{\prime}\right)dx^{\prime}+\frac{1}{2\pi j\omega\varepsilon_{0}}\intop_{x^{\prime}}\left\{ \frac{\partial^{2}}{\partial x^{2}}\left[K_{0}\left(\alpha\right)\right]J_{\mathrm{x}}\left(x^{\prime}\right)+j\beta\frac{\partial}{\partial x}\left[K_{0}\left(\alpha\right)\right]J_{\mathrm{z}}\left(x^{\prime}\right)\right\} dx^{\prime}
\end{equation}

\end{singlespace}

\begin{equation}
E_{\mathrm{y}}\left(x,y\right)=\frac{1}{2\pi j\omega\varepsilon_{0}}\frac{\partial}{\partial y}\intop_{x^{\prime}}\left\{ \frac{\partial}{\partial x}\left[K_{0}\left(\alpha\right)\right]J_{\mathrm{x}}\left(x^{\prime}\right)+j\beta K_{0}\left(\alpha\right)J_{\mathrm{z}}\left(x^{\prime}\right)\right\} dx^{\prime}
\end{equation}

\begin{singlespace}
\begin{equation}
E_{\mathrm{z}}\left(x,y\right)=\frac{k_{0}^{2}}{2\pi j\omega\varepsilon_{0}}\intop_{x^{\prime}}K_{0}\left(\alpha\right)J_{\mathrm{z}}\left(x^{\prime}\right)dx^{\prime}+\frac{1}{2\pi j\omega\varepsilon_{0}}\intop_{x^{\prime}}\left\{ j\beta\frac{\partial}{\partial x}\left[K_{0}\left(\alpha\right)\right]J_{\mathrm{x}}\left(x^{\prime}\right)-\beta^{2}K_{0}\left(\alpha\right)J_{\mathrm{z}}\left(x^{\prime}\right)\right\} dx^{\prime}
\end{equation}

\end{singlespace}

in which 
\begin{equation}
\alpha=\sqrt{q_{\mathrm{z}}^{2}-k^{2}}\sqrt{\left(x-x^{\prime}\right)^{2}+\left(y-b\right)^{2}}.
\end{equation}

The currents ($\mathrm{Re}\left(J_{\mathrm{z}}\left(x\right)\right)$) and
field distributions ($\left|E_{\mathrm{z}}^{2}\left(x,y\right)\right|$) associated
with the second and third SB modes ($2^{\prime}$  and  $3^{\prime}$ in
Fig. 6) at different frequencies are given in Figs. 8 and 9, respectively.

Two assumptions are made in the calculation of the surface modes.
The conductivity distribution based on the electrostatic charge distribution
is assumed to be only slightly perturbed by the modal fields, i.e.,
$\frac{\nabla.\,\mathbf{J}}{j\omega}\ll\rho$ where $\rho$ is the static
charge density and $\mathbf{J}$ is the modal current density. The
second assumption is that the the ground plane (and its ridge) are
far enough from the surface that the ground plane does not interact
with the (tightly-confined) modal fields. The parameters of the geometry
were chosen so that these assumptions are both valid. 

\bibliography{achemso}

\end{document}